\documentstyle[12pt]{article}
\begin{document}
\mbox{~}

\begin{center}
{\Large\bf Proper time and Minkowski structure\\[0.3cm]
           on causal graphs} \\[2cm]
{\Large Thomas Filk }\\[0.5cm]
{\large Institute for Theoretical Physics \\
Universit\"at Freiburg \\
Hermann-Herder-Str.\,3\\
D-79094 Freiburg \\
Germany}\\[0.2cm]
E-mail: \verb+thomas.filk@t-online.de+\\
~ 
\end{center}
\vspace{1.2cm}

\thispagestyle{empty}
\setcounter{page}{0}
\begin{abstract}
For causal graphs we propose a
definition of proper time which for small scales is based on the
concept of volume, while for large scales the usual definition of
length is appied. The scale where the change
from ``volume'' to ``length'' occurs is related to the size of
a dynamical clock and defines a natural cut-off for this type of
clock. By changing the cut-off volume we may probe the geometry of
the causal graph on different scales and thereby define a continuum
limit. This provides an alternative to the standard coarse graining
procedures.
For regular causal lattices (like e.g.\ the 2-dim.\ light-cone
lattice) this concept can be proven to lead to a Minkowski structure.
An illustrative example of this approach is provided by the
breather solutions of the Sine-Gordon model on a 2-dimensional
light-cone lattice.
\end{abstract}

\vspace{2cm}

\mbox{~} \hfill
\begin{minipage}[t]{6cm}
\begin{center}
University of Freiburg \\
February 2001\\
preprint THEP 01/03
\end{center}
\end{minipage}
\newpage

\section{Introduction}
Recently, there has been an increased interest in discrete causal 
structures as models for spacetime at small scales
\cite{Bombelli,Brightwell,Sorkin,Reid,Requardt,Loll}. 
In some cases, the dynamics which generates
these spacetime structures assigns to events a 
time-labeling which is unphysical in the sense 
that it fixes equal time relations and rather resembles a Newtonian
spacetime than a Minkowskian one. This time-labeling, however, is not
expected to be related to the physical time which an intrinsic observer 
would experience and which should exhibit a Minkowski or 
Lorentz structure. (In \cite{Sorkin2} an axiom of
``discrete general covariance'' requires that physical quantities
do not depend on this labeling.)

To recall an expression of Einstein, the physical time is 
``the time you can read from a clock''. A clock is a physical system
which takes part in the dynamics and which allows to identify and
count a characteristic time scale. In the ideal case the evolution 
of this system is periodic and the number of ``ticks'' between 
two events on
the world-line of this clock is equal to the length (the proper time) 
of this section of the world-line. Furthermore, a clock
should have a negligible influence on the spacetime structure and
should be small compared to the scale of geometrical variations of the
underlying geometry.
In order that different types of clocks define the same geometry
(up to a change of scale), the fundamental dynamics  
should be universal. In continuum physics this property is
guaranteed by local Lorentz invariance of the dynamics as
expressed, e.g., in the wave operator and the Dirac operator for
bosonic and fermionic fields, respectively. 

Unfortunately, discretized analogues of the wave operator
or the Dirac operator are only known for special cases: 
In \cite{Bobenko} the Sine-Gordon model is studied on a light-cone
lattice (see also sect.\ \ref{sec5}) and
Feynman gives a prescription to obtain the 2-dimensional Dirac
propagator from a random walk prescription \cite{Feynman} which is
related to the 6-vertex model in statistical mechanics \cite{Baxter}.
For unoriented graphs, which serve
as discretizations of euclidean spacetime, the situation is different
in that an analogue of the Laplacian is well known and extensively
studied in the literature (for a collection of results on the spectrum 
of the graph laplacian see e.g.\ \cite{Laplace}.)
For the same
reason, the ``summation over paths''-representation of propagators
or Green functions in Minkowski space is by far not as well
understood as in euclidean spaces (see e.g.\ \cite{Carlini}). 
It is known that a summation over
timelike paths will not lead to the correct expressions, non-physical
paths violating causality have to be taken into account. While the
physical paths formally are ``weighted'' by a phase, the unphysical
contributions are damped by real factors. However, this prescription
is derived from the formal expression
\begin{equation}
        G(x,y) ~=~ \sum_{{\rm paths}~ x\rightarrow y}
  \exp \left( {\rm i} \int {\rm d}\tau \sqrt{-\dot{x}^2}\right) \, , 
\end{equation}
and any attempts to give it a precise meaning which
also works for discretized versions (e.g.\ on causal graphs) have so
far failed. It should be emphasized, however, that the real dynamics
is not given by some particle or field which propagates {\em on}
an existing physical spacetime but the dynamics should be such that 
spacetime {\em and} matter are generated simultaneously (a possible
mechanism for this is indicated in \cite{Sorkin2}). 
 
If the microscopic dynamics of fields on a causal graph would be
known, we might be able to derive a definition of proper time using
the propagator function (see sect.\ \ref{sec3}, where we will also 
show why
the definition of proper time as ``the number of links'' of a timelike
path is inadequate for many cases). There is one property, however,
which any definition of proper time should satisfy if it is to lead
to a Lorentz structure: in flat Minkowski spacetime the geodesic
distance between any two events, $(a,b)$, should be in on-to-one 
correspondence to Alexandrov volumes $V[a,b]$
(i.e.\ the number of events $z$ such that $\{a_i<z<b_i\}$, where 
``$<$'' denotes the causal ordering of events). 
This defines an equivalence relation on the space of event
pairs, and although the volume is not the measure of proper time, 
it can be used to define such a measure (sect.\ \ref{sec2}). 

In general relativity we require that each event has a neighborhood
such that within this neighborhood special relativity is approximated
to a given accuracy (this excludes the singularities in the center of
black holes etc.). Especially, the world volume spanned by the 
spacial extend and the unit of time of clocks should be within
such a neighborhood. On the other hand, we expect the spacetime 
structure to become non-trivial again for very small scales and if
we assume spacetime to be discrete at Planck scale
it remains questionable, if clocks can be arbitrarily small.
Any dynamical clock has an intrinsic cut-off corresponding to
its characteristic time unit. If there exist clocks that can measure
time down to Planck scale and resolve the discreteness
of spacetime, then it is not obvious at all that these
clocks will define the same Minkowski structure as a clock for which
the smallest measurable time scale is within the region where the
approximation of spacetime by a flat Minkowski space is reasonable. 

We present a definition of proper time which associates to each
clock a characteristic Alexandrov volume $\gamma$ which sets the time
unit for this clock: Two timelike events $(a,b)$ are separated by 
a proper time of ``one tick'' of this clock, if the Alexandrov
volume corresponding to these two points is $\gamma$. Two timelike
events $(a,b)$ are separated by ``two ticks'', if there exists a
third event $c$ such that the Alexandrov volumes of $(a,c)$ and 
$(c,b)$ both are equal to $\gamma$ and if there is no other event $c'$,
such that $V[a,c']>V[a,c]$ and $V[c',b]>V[c,b]$ (for details see 
sect.\ \ref{sec4}). Different types of clocks will be associated with
different $\gamma$, but $\gamma$ should not be too large, as in this 
case the clock will not resolve the large scale fluctuations of 
spacetime geometry, and $\gamma$ should not be too small, as in this 
case the clock will resolve the discreteness of spacetime. Within 
these limits a change of $\gamma$ is merely a change of time 
scale and we probe the continuum theory. 
Two such clocks will define the same geometry, and we refer 
to such clocks as ``standard'' clocks. In principal, 
$\gamma$ can be arbitrary small (even $\gamma=0$, in which 
case we recover the proper time definition related to the  ``number 
of links''), but in this case even the large scale structure defined 
by such a clock may be different from the one defined by a standard 
clock.

This paper is organized as follows. In sect.\ \ref{sec2} we give 
some definitions related to causal graphs and causal sets and recall 
some well known relations between volume and proper time on flat 
Minkwoski spaces. In sect.\ \ref{sec3} we show that the definition 
of proper time as the number of links of a timelike path does 
not lead to a Minkowski structure on special causal graphs as, e.g.,
the light-cone lattice. An example will illustrate how the knowledge of 
the ``summation over path''-prescription might lead to a definition of
proper time. In sect.\ \ref{sec4} we present our definition of proper 
time which is based on Alexandrov sets and in sect.\ \ref{sec5} we
will use the breather solutions of the Sine-Gordon theory to illustrate 
the idea of dynamical clocks on causal graphs in a special case.
Some remarks will conclude this paper. 
 
\section{Definitions and preliminary remarks}
\label{sec2}

We start by giving some definitions related to causal graphs.

{\sc Def.:} A {\em simple directed graph} is a non-reflexive,
asymmetric relation $E$ on a set $V$, the set of vertices (or events).

Hence, if $(a,b)\in E$, then $a\neq b$ and $(b,a)\not\in E$.
We say that $a$ and $b$ are connected by a causal link (or edge). 
Instead of directed graphs one also speaks of oriented graphs.
Furthermore, two events cannot be connected by more than one link.
We will always assume $V$ to be countable.

{\sc Def.:} A {\em directed path} on a simple directed graph from 
vertex $a$ to vertex $b$ is a succesion of $N$ pairs 
$(c_i,c_{i+1})\in E$ ($i=0,...,N-1$) such that $c_0=a$ and $c_N=b$. 
This path is said to be of {\em length} $N$. If $a=b$ the path is
said to be {\em closed}.

{\sc Def.:} A {\em causal graph} is a simple directed graph such that
there are no closed directed paths.

A causal graph defines a causality relation ``$<$'': We write $a<b$, 
iff there 
exists a directed path from vertex $a$ to vertex $b$. This relation
is asymmetric, non-reflexive and transitive. Given a causality
relation, we can speak of timelike and spacelike 
events, the set of future events, and the set of past events, etc.

{\sc Def.:} For two vertices $a$ and $b$ we define the 
{\em Alexandrov set} to be $[a,b]=\{z|a<z<b\}$. The {\em volume}
of the Alexandrov set (the number of its elements) will be 
denoted by $V[a,b]$.
\newpage

In the following, we will require the property of local finiteness, 
i.e., for any pair of events the Alexandrov set is finite. Under 
these conditions a causal graph uniquely defines a causal set 
\cite{Sorkin}. 
It should be noted, however, that causal sets and causal graphs are
not equivalent. Although one can construct a unique causal graph 
from a causal set by ``deleting'' all relations which follow from
transitivity, this causal graph might be different from the graph one
has started with, as we do not require that the links in $E$ are free
of any transitivity relations. (There also seems to be a slight 
difference between causal sets and causal graphs related to the 
interpretation of the spacetime structure: For causal sets it is
sometimes assumed that the number of events which
are ``lightlike'' is of measure zero \cite{Brightwell}; this is not 
the case for causal graphs.)   

We now recall some well-known facts about the volume of Alexandrov
sets and its relation to proper time in standard $d$-dimensional
Minkowski space. We will use the convention that $(a-b)^2>0$ 
implies that $a$ and $b$ are timelike (and we will use units such
that $c=1$). $\tau(a,b)=\sqrt{(a-b)^2}$ will be the
proper time distance between $a$ and $b$.

If $(a_1,b_1)$ and $(a_2,b_2)$ are two pairs of timelike events 
in flat Minkowski space such that $\tau(a_1,b_1)=\tau(a_2,b_2)$, 
then the volumes of the corresponding 
Alexandrov sets are equal, $V[a_1,b_2]=V[a_2,b_2]$, and vice versa:
equal volume of the Alexandrov sets implies equal proper time
distance. This relation expresses the
fact that the Alexandrov volume as well as the proper time are
Lorentz invariants. 

Let $c$ and $d$ be two spacelike events (i.e.\ $(c-d)^2<0$), then
\[             (c-d)^2 ~=~ -\tau(a,b)^2    \, , \]
where $a$ and $b$ are two timelike events such that $c$ and $d$ are
in $[a,b]$ and such that there is no other
Alexandrov set with this property having a smaller volume. (In more
than two dimensions, $a$ and $b$ are not uniquely determined by $c$
and $d$ but different choices for $(a,b)$ will have the same proper 
time distance.) Hence, once we have determined the timelike distances
also the spacelike distances and the geometry are fixed. Therefore, 
in the following
we will only be concerned with the definition of proper time
and not with the determination of the full metrical structure.
Note, however, that the relations above do not hold in curved 
spacetime. 

Obviously, the volume of Alexandrov sets is not the {\em measure} 
of proper time (apart from ``spacetime'' dimension 1), as the relation 
between volume and time
is non-linear and dimension dependent. However, we can reconstruct the
proper time (up to an overall factor) from the knowledge of
the Alexandrov volumes. This procedure is independent of the dimension
and well known so that we only sketch the idea. Once we know how to
double or bisect a certain distance the rest follows from suitable 
iteration of these steps. 

For two timelike events $a$ and 
$b$ consider all events $c$ such that $V[a,c]=V[c,b]$. Among this set
choose $c$ such that $V[a,c]$ is maximal. Then $c$ lies on the geodesic
line connecting $a$ and $b$ and $\tau(a,c)=\tau(a,b)/2$. Similarly,
consider all events $d$ such that $V[a,b]=V[b,d]$ and choose $d$ in this
set such that $V[b,d]$ ist maximal, then $\tau(a,d)=2 \tau(a,b)$.

The {\em proper time distance} between 
two timelike events is equal to the maximal proper time length of a 
timelike path connecting these two events.

\section{Mathematical and ``propagator'' proper time\\ on causal graphs}
\label{sec3}

In this section, we will investigate two suggestive definitions of
proper time for causal graphs; one is the so-called mathematical
proper time related to the number of links of a directed path and the
other follows from a propagation prescription for a point particle
on a causal graph. Applied to the simple example of a 2-dimensional
light-cone lattice, both definitions fail to reproduce the 
usual Minkowski structure.  

{\sc Def.:} The {\em mathematical proper time} of a directed path
from event $a$ to event $b$ is equal to the number of links $N$, i.e.,
the length of this path. 
The {\em mathematical proper time distance} between two 
timelike events $a$ and $b$ is equal to the maximal length of a 
directed path from $a$ to $b$.

Let us now apply this definition to the 2-dimensional light-cone 
lattice (fig.1). All events on a fixed $t$-slice within the future 
light-cone have the same mathematical proper time distance from 
the event $a$. One might object that the graphical embedding of the
light-cone lattice in the
2-dimensional plane of the paper is irrelevant for the
intrinsic properties and misleading and that a suitable coordinate 
transformation will even make the equal time slices of 
standard Minkowski spacetime look horizontal. However, as we shall
see in sect.\ \ref{sec5}, the dynamical proper time as experienced by
an intrinsic observer and his clocks will show the familiar hyperbolic
curves and has nothing to do with the horizontal equal time slices
of the light-cone lattice.
\setlength{\unitlength}{0.8pt}

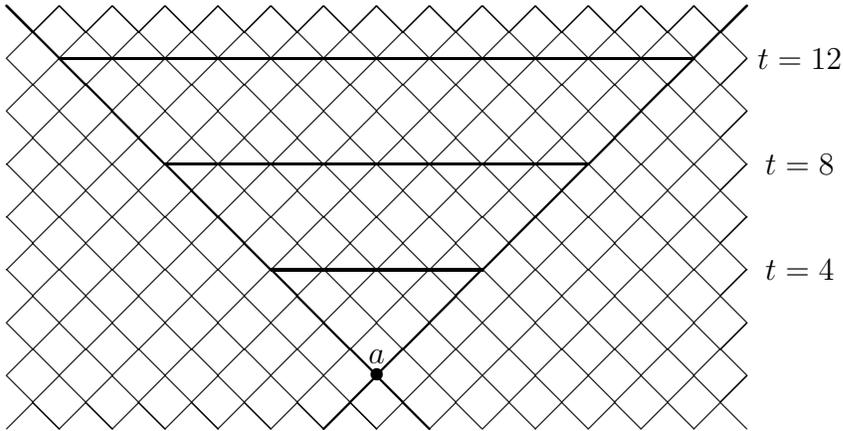
\begin{figure}[tbp]
\begin{picture}(400,200)(25,0)
\multiput(25,0)(25,0){7}{\line(1,1){200}}
\multiput(225,0)(25,0){7}{\line(-1,1){200}}
\put(25,25){\line(1,1){175}}
\put(25,50){\line(1,1){150}}
\put(25,75){\line(1,1){125}}
\put(25,100){\line(1,1){100}}
\put(25,125){\line(1,1){75}}
\put(25,150){\line(1,1){50}}
\put(25,175){\line(1,1){25}}
\put(25,25){\line(1,-1){25}}
\put(25,50){\line(1,-1){50}}
\put(25,75){\line(1,-1){75}}
\put(25,100){\line(1,-1){100}}
\put(25,125){\line(1,-1){125}}
\put(25,150){\line(1,-1){150}}
\put(25,175){\line(1,-1){175}}
\put(200,0){\line(1,1){175}}
\put(225,0){\line(1,1){150}}
\put(250,0){\line(1,1){125}}
\put(275,0){\line(1,1){100}}
\put(300,0){\line(1,1){75}}
\put(325,0){\line(1,1){50}}
\put(350,0){\line(1,1){25}}
\put(375,25){\line(-1,1){175}}
\put(375,50){\line(-1,1){150}}
\put(375,75){\line(-1,1){125}}
\put(375,100){\line(-1,1){100}}
\put(375,125){\line(-1,1){75}}
\put(375,150){\line(-1,1){50}}
\put(375,175){\line(-1,1){25}}
\thicklines
\put(175,0){\line(1,1){200}}
\put(225,0){\line(-1,1){200}}
\put(200,34){\makebox(0,0){$a$}}
\put(200,25){\makebox(0,0){$\bullet$}}
\put(50,175){\line(1,0){300}}
\put(100,125){\line(1,0){200}}
\put(150,75){\line(1,0){100}}
\put(400,175){\makebox(0,0){$t=12$}}
\put(400,125){\makebox(0,0){$t=8$}}
\put(400,75){\makebox(0,0){$t=4$}}
\end{picture}
\caption{\label{light}%
Equal time slices for the mathematical distance on the light-cone lattice}
\end{figure}

Next, we show how to derive a definition of proper time from a 
random walk process of a point-like particle on the causal graph. 
This process does not represent the dynamics of real physical 
particles, as in this case the paths are 
``weighted'' by phases instead of simple probabilities
and non-causal paths have to be taken into account. The
concept merely serves as an example how the dynamics of a 
particle may be used to define a proper time.

For two timelike events $a$ and $b$ let 
\begin{equation}
      G(a,b) ~=~ \sum_{{\rm paths~}a \rightarrow b}
                           {\rm e}^{-\mu L}  \, , 
\end{equation}
where $L$ is the length of the oriented path from $a$ to $b$.
The summation extends over all directed paths from $a$ to
$b$ and each path is weighted by a factor depending on its
length only. (This is reminiscent of the definition of the
euclidean Green function on undirected graphs.) We define:

Two pairs of events $(a_1,b_1)$ and $(a_2,b_2)$ have the same 
{\em propagator proper time distance}, iff $G(a_1,b_1)=G(a_2,b_2)$.

In complete analogy to the prescription sketched in the previous 
section for Alexandrov sets, we now can derive a proper time 
distance (up to an overall factor) from these ``Green functions''
$G(a,b)$. However, for the following argument we will only need the
equal time slices of this definition, i.e., for a given $a$ the
set of events $b$ for which $G(a,b)$ is constant. 

Let us consider again the light-cone lattice and let $a$ correspond 
to the point $(0,0)$. All paths from $(0,0)$ to a future
event $(x,t)$ have the same length $t$. Furthermore, the number of
paths from $(0,0)$ to $(x,t)$ is simply given by a binomial 
coefficient. We are interested in the asymptotic behavior and
make the usual gaussian approximation to obtain
\begin{equation}
       G((0,0),(x,t)) ~ \simeq ~ \frac{2^t}{t} 
      \exp \left( - 2 \frac{x^2}{t} - \mu t \right) \, .  
\end{equation}
An equal time slice is
given by the set of events for which $G$ is constant. The result
depends on the value of $\mu$: For $\mu<\ln 2$, the function $G$
increases with increasing $t$ and we ``almost'' obtain
a hyperbolic structure, but with logarithmic corrections and a
$t$-dependent center. For $\mu>\ln2$, vertices of equal time are
distributed along the section of a ``distorted'' circle which rather
resembles the euclidean case.

\section{Definition of proper time from Alexandrov\\ sets}
\label{sec4}

We now describe how to obtain the proper time of a directed path
using Alexandrov sets. 

We choose some reference volume $\gamma$ and define 
$\tau_\gamma(a,b)=1$, iff the Alexandrov volume $V[a,b]=\gamma$. 
Hence, $\gamma$ defines the unit of proper time. For the moment,
we will keep $\gamma$ fixed and postpone the discussion about
different and suitable choices of $\gamma$ to the second part of this 
section. In terms of physics, we associate $\gamma$ with a cut-off 
for a dynamical clock. $\gamma$ represents the world-volume of this
clock for one ``tick''. Such a ``$\gamma$-clock'' will not be able 
to resolve time on a scale smaller than $\tau_\gamma=1$. 

Now, let $C$ be a directed path from $x$ to $y$ and 
$V[x,y]\gg \gamma$. The proper time of $C$ between the events $x$ 
and $y$ is determined as follows: Choose events $z_0,z_1,...,z_N$ on
$C$ such that (i) $z_0=x$, (ii) $V[z_N,y]< \gamma$ and
(iii) $V[z_i,z_{i+1}]=\gamma$. If such events can be found we say that
the proper time (in units defined by $\gamma$) of the section of $C$ 
between $x$ and $y$ is between $N$ and $N+1$. A better resolution 
cannot be achieved with this type of clock. (For technical or
numerical purposes it might be useful to define some interpolating
procedure, but this will not be done here.)

In general, it will not be possible to find points $z_i$ such that the
conditions above are satisfied exactly and we can only require that
$V[z_i,z_{i+1}] \approx \gamma$ ``as good as possible''. This 
statement can be made more precise by formulating a variational
principle for the choices of $\{z_i\}$. Replace (iii) above by the
condition that
\begin{equation}
   S ~=~ \sum_{i=0}^{N-1} \left( V[z_i,z_{i+1}] - \gamma \right)^2 
\end{equation}
is minimal. 

This assignment of a proper time to a directed path is approximative 
in at least two respects. First, if $V[x,y]<\gamma$, it does not make 
sense to assign a precise proper time distance to $x$ and $y$. 
In this case we only may say that the distance (when measured with a
$\gamma$-clock) is smaller than 1. Second, even if the proper time
of a section of $C$ is large, we can only determine its value
up to one unit of this $\gamma$-clock. Both properties are to be
expected when time ``is read from a clock''.

Let us now see how the given definition of proper time works for 
the two-dimensional light-cone lattice. Figure 2 shows four points 
having equal proper time distance from $(0,0)$ for $\gamma=34$ 
(which is very small, cf.\ the numbers given below). The 
approximation of the hyperbolic structure of an equal-time slice is 
better for larger values of $\gamma$. Problems seem to occur when we
approach the light-cone, but this is to be expected: If a clock moves
relativ to an observer with a velocity such that its spacial extend 
by Lorentz contraction approaches the size of the lattice spacing 
(for instance Planck length) we expect to see deviations from 
continuum physics. 
\setlength{\unitlength}{0.7pt}

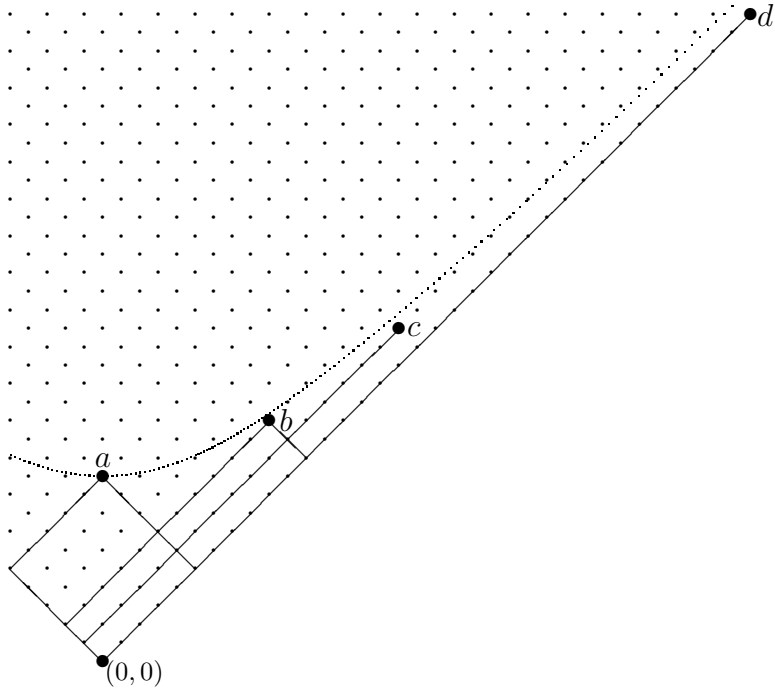
\begin{figure}[tbp]
\begin{picture}(400,400)(-50,-10)
\put(50,0){\makebox(0,0){$\cdot$}}
\put(40,10){\makebox(0,0){$\cdot$}}
\put(60,10){\makebox(0,0){$\cdot$}}
\put(30,20){\makebox(0,0){$\cdot$}}
\put(50,20){\makebox(0,0){$\cdot$}}
\put(70,20){\makebox(0,0){$\cdot$}}
\multiput(20,30)(20,0){4}{\makebox(0,0){$\cdot$}}
\multiput(10,40)(20,0){5}{\makebox(0,0){$\cdot$}}
\multiput(0,50)(20,0){6}{\makebox(0,0){$\cdot$}}
\multiput(10,60)(20,0){6}{\makebox(0,0){$\cdot$}}
\multiput(0,70)(20,0){7}{\makebox(0,0){$\cdot$}}
\multiput(10,80)(20,0){7}{\makebox(0,0){$\cdot$}}
\multiput(0,90)(20,0){8}{\makebox(0,0){$\cdot$}}
\multiput(10,100)(20,0){8}{\makebox(0,0){$\cdot$}}
\multiput(0,110)(20,0){9}{\makebox(0,0){$\cdot$}}
\multiput(10,120)(20,0){9}{\makebox(0,0){$\cdot$}}
\multiput(0,130)(20,0){10}{\makebox(0,0){$\cdot$}}
\multiput(10,140)(20,0){10}{\makebox(0,0){$\cdot$}}
\multiput(0,150)(20,0){11}{\makebox(0,0){$\cdot$}}
\multiput(10,160)(20,0){11}{\makebox(0,0){$\cdot$}}
\multiput(0,170)(20,0){12}{\makebox(0,0){$\cdot$}}
\multiput(10,180)(20,0){12}{\makebox(0,0){$\cdot$}}
\multiput(0,190)(20,0){13}{\makebox(0,0){$\cdot$}}
\multiput(10,200)(20,0){13}{\makebox(0,0){$\cdot$}}
\multiput(0,210)(20,0){14}{\makebox(0,0){$\cdot$}}
\multiput(10,220)(20,0){14}{\makebox(0,0){$\cdot$}}
\multiput(0,230)(20,0){15}{\makebox(0,0){$\cdot$}}
\multiput(10,240)(20,0){15}{\makebox(0,0){$\cdot$}}
\multiput(0,250)(20,0){16}{\makebox(0,0){$\cdot$}}
\multiput(10,260)(20,0){16}{\makebox(0,0){$\cdot$}}
\multiput(0,270)(20,0){17}{\makebox(0,0){$\cdot$}}
\multiput(10,280)(20,0){17}{\makebox(0,0){$\cdot$}}
\multiput(0,290)(20,0){18}{\makebox(0,0){$\cdot$}}
\multiput(10,300)(20,0){18}{\makebox(0,0){$\cdot$}}
\multiput(0,310)(20,0){19}{\makebox(0,0){$\cdot$}}
\multiput(10,320)(20,0){19}{\makebox(0,0){$\cdot$}}
\multiput(0,330)(20,0){20}{\makebox(0,0){$\cdot$}}
\multiput(10,340)(20,0){20}{\makebox(0,0){$\cdot$}}
\multiput(0,350)(20,0){21}{\makebox(0,0){$\cdot$}}
\put(50,0){\makebox(0,0){$\bullet$}}
\put(50,100){\makebox(0,0){$\bullet$}}
\put(140,130){\makebox(0,0){$\bullet$}}
\put(210,180){\makebox(0,0){$\bullet$}}
\put(400,350){\makebox(0,0){$\bullet$}}
\put(50,0){\line(1,1){350}}
\put(0,50){\line(1,1){50}}
\put(50,0){\line(-1,1){50}}
\put(100,50){\line(-1,1){50}}
\put(30,20){\line(1,1){110}}
\put(160,110){\line(-1,1){20}}
\put(40,10){\line(1,1){170}}
\put(220,170){\line(-1,1){10}}
\put(67,-6){\makebox(0,0){{\footnotesize $(0,0)$}}}
\put(50,109){\makebox(0,0){$a$}}
\put(149,131){\makebox(0,0){$b$}}
\put(218,180){\makebox(0,0){$c$}}
\put(408,350){\makebox(0,0){$d$}}
%
%
\qbezier[40](0,111.8)(50,89)(100,111.8)
\qbezier[100](100,111.8)(170,140)(390,354.7)
\end{picture}
\caption{\label{light}%
Alexandrov sets of equal ``volume'' $\gamma=34$. The dotted 
curve represents the Minkowski equal time slice.}
\end{figure}

A more severe obstacle seems to be that for finite
$\gamma$ there will be an event {\em on} the light-cone for which
the Alexandrov volume is equal to $\gamma$ (event $d$ in fig.\ 2). 
Hence, paths along the
light-cone have a non-vanishing proper time. The light-cone 
structure seems to be a general problem for models with discretized 
spacetime. 

But before we discard this approach completely let us put in
some more realistic numbers. Assume the lattice scale to be 
equal to the Planck scale ($l_P\approx 10^{-33}$\,cm 
and $t_P\approx 10^{-44}$\,sec). At present we probe
physics below 1\,TeV$\approx 10^{-17}$\,cm. (That does not imply that
we can measure space and time with this accuracy.) 
This is 16 orders of magnitude above Planck scale. A clock 
(in 4 dimensional spacetime) with this precision would correspond 
to a value of $\gamma=10^{64}$. The associated linear distance
on the light-cone would be $10^{64}\cdot l_P\approx 10^{31}$\,cm. 
This is several orders of magnitude larger than the present radius
of the universe. So we are still far away from observing deviations 
from continuum physics. (However, an increase of 3 orders of magnitude 
in energy would correspond to a linear distance on the 
light-cone of about $10^{19}$\,cm or 10\,parsec, which can be
measured directly by using the annual parallax of the earth. 
Maybe we are not so far away from seeing Planck scale physics.)

We might also have argued that this is just an unphysical artifact
of our prescription and that a better understanding of the dynamics
related to causal graphs will solve this problem. In any case,
physical clocks cannot be accelerated to travel {\em on} the
light cone.   

Finally, we want to address the question of how to choose proper 
values for $\gamma$. In principle, $\gamma$ may be any non-negative 
integer starting from 0
(in which case the above definition of proper time reduces to the
mathematical proper time -- the ``number of links''). Therefore,
$\gamma$ labels an infinite family of definitions of proper time.
The natural question is if these families give rise to the same
geometry (up to an overall factor). In general, this will not be
the case. If $\gamma$ is too small, artifacts of
the discreteness of spacetime may have an influence on the large scale 
structure, as we have seen in our discussion of the light-cone
problem. For numerical purposes $\gamma$ should be larger than the
mathematical diameter (the length of the longest directed path) of 
the causal graph. 

$\gamma$ should not be too large either, as in this case
the large scale geometry is washed out. Macroscopic curvature effects,
i.e.\ the realm of classical (non-quantum) general relativity,
should be negligible inside a world-volume of size $\gamma$. 
A similar requirement has to be made for all time measuring 
instruments in general relativity. Hence, apart from regions close to
singularities (e.g.\ inside of black holes), this leaves a large range
for $\gamma$ in our world.

On the other hand, if for a causal graph we find a range of $\gamma$
such that within this range a change of $\gamma$ defines the same
geometry (up to a scale), we may say that this range probes the
continuum limit of the graph.

Up to now we have treated the world-volume of a clock during one
period as an Alexandrov set. In physical applications (or for the
breather solution which we will discuss in the next section), the
actual world-volume might have a different shape. In any case, this
volume is Lorentz invariant and in one-to-one correspondence to 
proper time. We have chosen Alexandrov sets just for convenience.

\section{Breathers in the discretized Sine-Gordon\\ 
model as an example for dynamical clocks}
\label{sec5}

We now want to illustrate the concepts defined in the previous
section by a simple example: the Sine-Gordon model on a 2-dim.\ 
light-cone lattice. This model has been studied in \cite{Bobenko}.
The Sine-Gordon theory in two dimensions is
known to have periodic solutions - the so-called breather solutions 
\cite{Soliton} - which may serve as dynamical clocks. 

The field equation of the Sine-Gordon model is:
\begin{equation}
\label{cont}
  \frac{\partial^2 \varphi}{\partial t^2} -         
    \frac{\partial^2 \varphi}{\partial x^2} + g \sin \varphi ~=~ 0 \;, 
\end{equation}             
where $\varphi$ denotes an angular variable. 

We now discretize the above equation for a light-cone lattice 
(for the labeling of points cf.\ fig.\,\ref{light}):
\begin{eqnarray}
\label{discrete}
 0 &=& \varphi(m+1,n+1) + \varphi(m,n) - 
                                  \varphi(m,n+1) - \varphi(m+1,n) + \\
\nonumber   
 &  & \frac{1}{4}g \left[
    \sin \varphi(m,n)+\sin \varphi(m\!+\!1,n)+\sin \varphi(m,n\!+\!1)     
     +  \sin \varphi(m\!+\!1,n\!+\!1)  \right]  \;. 
\end{eqnarray}    

\begin{figure}[htbp]
\begin{picture}(200,240)(-100,-20)
\put(100,0){\makebox(0,0){$\bullet$}}
\put(100,100){\makebox(0,0){$\bullet$}}
\put(100,200){\makebox(0,0){$\bullet$}}
\put(50,150){\makebox(0,0){$\bullet$}}
\put(50,50){\makebox(0,0){$\bullet$}}
\put(150,150){\makebox(0,0){$\bullet$}}
\put(150,50){\makebox(0,0){$\bullet$}}
\put(200,100){\makebox(0,0){$\bullet$}}
\put(0,100){\makebox(0,0){$\bullet$}}
\put(100,0){\line(1,1){100}}
\put(100,0){\line(-1,1){100}}
\put(0,100){\line(1,1){100}}
\put(200,100){\line(-1,1){100}}
\put(50,50){\line(1,1){100}}
\put(150,50){\line(-1,1){100}}
\put(100,-10){\makebox(0,0){${\scriptstyle (m-1,n-1)}$}}
\put(121,100){\makebox(0,0){${\scriptstyle (m,n)}$}}
\put(100,210){\makebox(0,0){${\scriptstyle (m+1,n+1)}$}}
\put(176,49){\makebox(0,0){${\scriptstyle (m,n-1)}$}}
\put(176,151){\makebox(0,0){${\scriptstyle (m+1,n)}$}}
\put(235,100){\makebox(0,0){${\scriptstyle (m+1,n-1)}$}}
\put(24,49){\makebox(0,0){${\scriptstyle (m-1,n)}$}}
\put(24,151){\makebox(0,0){${\scriptstyle (m,n+1)}$}}
\put(-35,100){\makebox(0,0){${\scriptstyle (m-1,n+1)}$}}
\thicklines
\put(128,28){\vector(1,1){0}}
\put(178,78){\vector(1,1){0}}
\put(72,28){\vector(-1,1){0}}
\put(22,78){\vector(-1,1){0}}
\put(28,128){\vector(1,1){0}}
\put(78,178){\vector(1,1){0}}
\put(172,128){\vector(-1,1){0}}
\put(122,178){\vector(-1,1){0}}
\put(78,78){\vector(1,1){0}}
\put(128,128){\vector(1,1){0}}
\put(122,78){\vector(-1,1){0}}
\put(72,128){\vector(-1,1){0}}
\end{picture}
\caption{\label{light}%
Labeling of the vertices on a 2-dimensional light-cone lattice}
\end{figure}
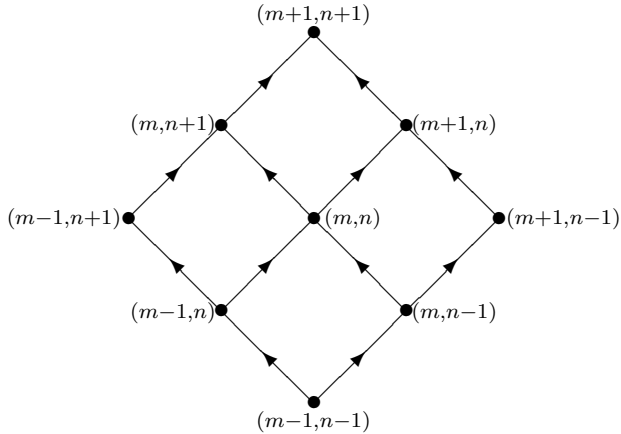

We are looking for solutions for which the values of the fields on
neighbored vertices do not differ much. Such weakly fluctuating 
solutions are obtained for $g\ll 1$. In this case the solutions
will be given in good approximation by the continuum equations.

We have labeled the points of the light-cone lattice by 
$(m,n)\simeq (t-x,t+x)$ which refers to a fixed background
spacetime with a Newtonian character. We know, however, that the
continuum equations are invariant under Lorentz transformations. This
invariance is not an invariance of the spacetime structure defined
by the lattice, but a property of the set of solutions of the equation:
If $\varphi(x,t)$ is a solution of eq.~(\ref{cont}), then 
\begin{equation}
\label{inv}
   \hat{\varphi}(x,t) ~=~
   \varphi\left(\,\gamma(v)(x-vt)\:,\:
                \gamma(v) (t-  v x)\, \right)
   ~~~{\rm with}~~ \gamma(v) = \frac{1}{\sqrt{1-v^2}}
\end{equation}
is also a solution of this equation for arbitrary $-1<v<1$. As long as
$g\gamma(v)\ll 1$,
the solutions of the discretized equation (\ref{discrete}) will be
approximated by the solutions of the continuum equation. Within the
framework of this
approximation the set of discretized solutions will share the same
invariance property (\ref{inv}). Hence, {\em not the structure or 
labeling of the lattice but the invariance of the space of solutions 
of the equations of motion - the dynamics - will define ``intrinsic 
spacetime''.}

The breather solutions of the Sine-Gordon equation may be interpreted
as metastable bound states of a soliton and an anti-soliton. At rest 
this solution is given by
\begin{equation}
\label{breather}
  \varphi(x,t) ~=~ - 4 \tan^{-1} \left[ \frac{a}{\sqrt{1-a^2}}
  \frac{\sin \sqrt{g(1-a^2)}(t-t_0)}{\cosh a \sqrt{g}(x-x_0)} \right]\;.
\end{equation}
The parameter $a$ has to satisfy the condition $0<a^2<1$ but is arbitrary
otherwise. The period for this solution is
\begin{equation}
\label{time}
     \Delta T_0 ~=~ \frac{2\pi}{\sqrt{g(1-a^2)}}~.
\end{equation}
In order to employ the breather solution as a dynamical clock we
have to choose a definite value for $a$, thereby setting a
time scale. A possible choice corresponds to the
case where $\varphi$ varies between $-\pi$ and $+\pi$, i.e.\
$a^2=1/2$ or $\Delta T_0=\sqrt{8\pi^2/g}$.

We may use the invariance of the field equation to obtain a breather
solution which moves with a velocity $v$. Not surprisingly, the
period as seen by an external observer (like us) changes to
\begin{equation}
\label{dilation}
      \Delta T_v ~=~ \frac{1}{\sqrt{1-v^2}} \Delta T_0 \;.
\end{equation}
The period of a moving bound state is longer by a factor of $\gamma(v)$
with respect to the period of a breather solution at rest. On the
other hand, the width of the breather solution is 
Lorentz contracted such that the world-volume
during one period (defined, e.g., by the set of events for which
$|\varphi|$ is larger than some constant) is unchanged.
This world volume does not resemble an Alexandrov set at all 
(cf.\ fig.\,4), but it is in one-to-one correspondence
to proper time. Measuring time means counting elementary volumes.
\setlength{\unitlength}{0.7cm}

\begin{figure}[hbtp]
\begin{picture}(7,8)(-5,0)
\put(0,0){\line(1,1){3.142}}
\put(0,0){\line(-1,1){3.142}}
\put(0,0){\line(0,1){6.283}}
\put(0,6.283){\line(1,-1){3.142}}
\put(0,6.283){\line(-1,-1){3.142}}
\put(2.5,0.5){\makebox(0,0){$(a)$}}
\put(-0.83,0.6){\line(1,0){1.66}}
\put(-1.18,0.8){\line(1,0){2.36}}
\put(-1.37,1.0){\line(1,0){2.74}}
\put(-1.48,1.2){\line(1,0){2.96}}
\put(-1.55,1.4){\line(1,0){3.10}}
\put(-1.56,1.6){\line(1,0){3.12}}
\put(-1.53,1.8){\line(1,0){3.06}}
\put(-1.46,2.0){\line(1,0){2.92}}
\put(-1.32,2.2){\line(1,0){2.64}}
\put(-1.11,2.4){\line(1,0){2.22}}
\put(-0.70,2.6){\line(1,0){1.40}}
\qbezier(-1.18,3.942)(-1.96,4.712)(-1.18,5.484)
\qbezier(1.18,3.942)(1.96,4.712)(1.18,5.484)
\qbezier(-1.18,3.942)(0,3.182)(1.18,3.942)
\qbezier(-1.18,5.484)(0,6.252)(1.18,5.484)
\qbezier(-1.18,0.8)(-1.96,1.57)(-1.18,2.342)
\qbezier(1.18,0.8)(1.96,1.57)(1.18,2.342)
\qbezier(-1.18,0.8)(0,0.04)(1.18,0.8)
\qbezier(-1.18,2.342)(0,3.11)(1.18,2.342)
\put(-0.83,3.742){\line(1,0){1.66}}
\put(-1.18,3.942){\line(1,0){2.36}}
\put(-1.37,4.142){\line(1,0){2.74}}
\put(-1.48,4.342){\line(1,0){2.96}}
\put(-1.55,4.542){\line(1,0){3.10}}
\put(-1.56,4.742){\line(1,0){3.12}}
\put(-1.53,4.942){\line(1,0){3.06}}
\put(-1.46,5.142){\line(1,0){2.92}}
\put(-1.32,5.342){\line(1,0){2.64}}
\put(-1.11,5.542){\line(1,0){2.22}}
\put(-0.70,5.742){\line(1,0){1.40}}
\end{picture}
\begin{picture}(7,8)(-6,0)
\put(0,0){\line(1,1){5.4426}}
\put(0,0){\line(-1,1){1.8134}}
\put(0,0){\line(1,2){3.6276}}
\put(3.6276,7.2552){\line(1,-1){1.8134}}
\put(3.6276,7.2552){\line(-1,-1){5.4426}}
\put(3.0,0.5){\makebox(0,0){$(b)$}}
\qbezier(-0.73,1.2)(0.15,2.5)(1.68,3.25)
\qbezier(0.08,0.4)(1.5,1.1)(2.53,2.4)
\qbezier(1.68,3.25)(2.8,3.7)(2.84,3.2)
\qbezier(2.84,3.2)(2.83,2.83)(2.53,2.4)
\qbezier(-0.73,1.2)(-1.3,0.45)(-0.81,0.2)
\qbezier(-0.81,0.2)(-0.3,0.2)(0.08,0.4)
\put(-0.81,0.2){\line(1,0){0.15}}
\put(-1.02,0.4){\line(1,0){1.10}}
\put(-1.03,0.6){\line(1,0){1.51}}
\put(-0.97,0.8){\line(1,0){1.78}}
\put(-0.85,1.0){\line(1,0){1.95}}
\put(-0.73,1.2){\line(1,0){2.08}}
\put(-0.58,1.4){\line(1,0){2.19}}
\put(-0.42,1.6){\line(1,0){2.25}}
\put(-0.25,1.8){\line(1,0){2.28}}
\put(-0.07,2.0){\line(1,0){2.29}}
\put(0.14,2.2){\line(1,0){2.24}}
\put(0.37,2.4){\line(1,0){2.16}}
\put(0.64,2.6){\line(1,0){2.04}}
\put(0.91,2.8){\line(1,0){1.85}}
\put(1.24,3.0){\line(1,0){1.59}}
\put(1.62,3.2){\line(1,0){1.22}}
\put(2.17,3.4){\line(1,0){0.56}}
\qbezier(1.084,4.828)(1.964,6.128)(3.494,6.878)
\qbezier(1.894,4.028)(3.314,4.728)(4.344,6.028)
\qbezier(3.494,6.878)(4.614,7.328)(4.654,6.828)
\qbezier(4.654,6.828)(4.644,6.458)(4.344,6.028)
\qbezier(1.084,4.828)(0.514,4.078)(1.004,3.828)
\qbezier(1.004,3.828)(1.514,3.828)(1.894,4.028)
\put(1.004,3.828){\line(1,0){0.15}}
\put(0.794,4.028){\line(1,0){1.10}}
\put(0.784,4.228){\line(1,0){1.51}}
\put(0.844,4.428){\line(1,0){1.78}}
\put(0.964,4.628){\line(1,0){1.95}}
\put(1.084,4.828){\line(1,0){2.08}}
\put(1.234,5.028){\line(1,0){2.19}}
\put(1.394,5.228){\line(1,0){2.25}}
\put(1.564,5.428){\line(1,0){2.28}}
\put(1.744,5.628){\line(1,0){2.29}}
\put(1.954,5.828){\line(1,0){2.24}}
\put(2.184,6.028){\line(1,0){2.16}}
\put(2.454,6.228){\line(1,0){2.04}}
\put(2.724,6.428){\line(1,0){1.85}}
\put(3.054,6.628){\line(1,0){1.59}}
\put(3.434,6.828){\line(1,0){1.22}}
\put(3.984,7.028){\line(1,0){0.56}}
\end{picture}
\caption{%
World-volumes of the breather solution for (a) $v=0$ and (b) $v=0.5$
and the corresponding Alexandrov sets. (The shaded areas represent 
events for which $|\varphi|>1.5$.)}
\end{figure}
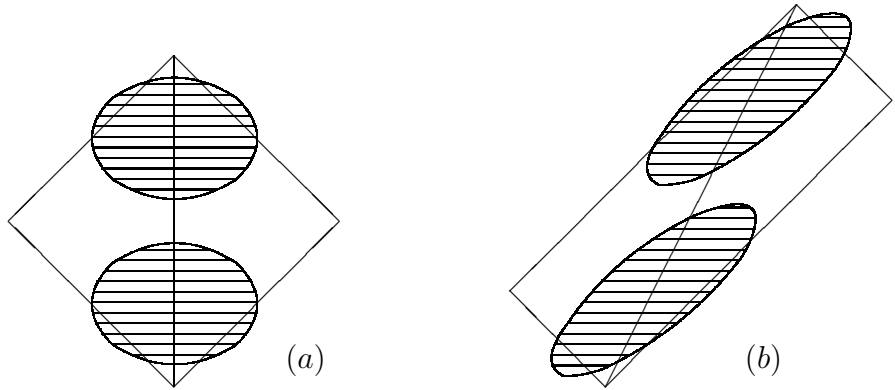

Intrinsically, the breather solution defines the proper time
distance between two timelike events. This structure is obviously
the Minkowski structure (as long as we do not approach the light-cone 
too closely).

\section{Concluding remarks}

We presented a definition of proper time for causal graphs which is 
based on the idea of dynamical clocks with a certain accuracy. We
have seen that
\begin{itemize}
\item[-]
the assignment of a proper time to a section of a directed path 
depends on the world volume $\gamma$ of a clock during one ``tick''
and is approximative because the resolution of proper time has a
natural cut-off depending on the type of clock.
\item[-]
the geometry of the causal graph may depend on this resolution. 
If the accuracy of the clock can resolve the discrete nature
of the causal graph, the large scale structure deviates from that
of a Minkowski space. In the other extreme the clock may smear out
large scale curvature dependencies. Only if the one-tick world volume
of the clock lies within the range where a Lorentz approximation to
spacetime is valid, the reconstructed geometry corresponds to the
one of general relativity. 
\item[-]
the breather solution of the Sine-Gordon equation is
an example of a clock moving on the light-cone lattice (which is
a causal graph). However, as emphasized before, the full theory should
generate spacetime and propagating fields simultaneously.
\end{itemize}
A few remarks are in order:
\begin{itemize}
\item[-]
Although we have used the language of causal graphs, the concept
may be applied to any discrete structure with a causality relation.
It might be especially well suited for causal sets.
\item[-]
One might wonder why the causal structure is not sufficient for a 
Minkowski or Lorentz structure to arise at large scales, which
is what many 
theorems on causal spaces imply \cite{Zeeman,Kronheimer,Malament}. 
(The notion of ``volume'' - number of events - should fix the metric
even completely.) However, in most cases at least continuity or the 
concept of parallel lines is required to prove these 
theorems. 
\item[-]
Apart from the physically motivated definition of proper time,
this mechanism might also be an alternative way to define 
coarse graining for causal graphs and causal sets, which is important 
for the definition of a continuum limit. 
In our approach, the assignment of proper times implies 
the continuum limit, as larger values of $\gamma$ 
probe larger scales of the causal graph. 

It would be interesting to 
see, if there is a relation (or even equivalence) to the coarse 
graining procedure proposed 
by the authors of \cite{Sorkin3}. They construct a coarse grained 
causal set from a random selection of events, which resembles 
the ``decimation'' in the context of spin systems. The approach 
presented here rather
resembles a block spin transformation where an effective magnetization
is assigned to a certain volume which then represents a point in the
coarse grained lattice. However, we never select specific volumes
to obtain a coarse grained lattice (which might violate Lorentz
invariance) but rather leave the causal graph unchanged and only
probe it on different scales. This procedure is closer to what we
actually do in Nature. 
\end{itemize}

\end{document}